\documentclass[showpacs,preprintnumbers,amsmath,amssymb,amsthm]{elsarticle}

\newcommand\etal{\mbox{\textit{et al. }}}

\usepackage{graphicx}
\usepackage{dcolumn}
\usepackage{bm}
\usepackage{txfonts}
\usepackage{subfigure}
\usepackage[dvips]{psfrag}
\usepackage{float}
\usepackage{color}

\graphicspath{{../FIGURAS7/}}

\begin{document}
\title{Viscous pumping inspired by flexible propulsion}

\author{Roger M. Arco, J. Rodrigo V\'elez-Cordero}
\address{Instituto de Investigaciones en Materiales, Universidad Nacional Aut\'onoma de
M\'exico, Apdo. Postal 70-360 M\'exico D.F. 04510, Mexico}
\author{Eric Lauga}
\address{Department of Applied Mathematics and Theoretical Physics,  University of Cambridge,
Centre for Mathematical Sciences, Wilberforce Road, Cambridge, CB3 0WA, United Kingdom}
\ead{e.lauga@damtp.cam.ac.uk}
\author{Roberto Zenit}
\address{Instituto de Investigaciones en Materiales, Universidad Nacional Aut\'onoma de M\'exico, Apdo. Postal 70-360
M\'exico D.F. 04510, Mexico}
\ead{zenit@unam.mx}

\date{\today}

\begin{abstract}
Fluid-suspended microorganisms have evolved different swimming and feeding strategies in order to cope with an environment dominated by viscous effects.   For instance ciliated organisms rely on the collective motion of flexible appendices to move and feed. By performing a non-reciprocal motion,  flexible filaments can produce a net propulsive force, or pump fluid, in the absence of inertia. Inspired by such fundamental concept, we propose a strategy to produce macroscopic pumping and mixing in creeping flow. We measure experimentally  the net motion of a Newtonian viscous fluid induced   by the reciprocal motion of a  flapper. When the flapper is rigid no net motion  is induced. In contrast, when the flapper is made of a flexible material, a net fluid pumping is measured. We quantify the effectiveness of this pumping strategy  and show that optimal pumping is  achieved when the length of the flapper is on the same order as the  elasto-hydrodynamic penetration length. We finally discuss the possible applications of flexible impellers in mixing operations at low Reynolds numbers.

\end{abstract}


\maketitle

\section{Introduction}

Fluid transport at high Reynolds numbers is a process particularly common in our daily experience. For example,  dispersion of a solute in water does not demand long times or large amounts of power per unit volume and effective mixing can be achieved by the simple reciprocal (time-reversible) motion of a rigid impeller such as a spoon. Even though the impeller may have a small contact area compared with the volume of
fluid requiring stirring, mixing is greatly enhanced by the large-scale, convective motion resulting from inertial effects in the flow. In particular, such  motion can be observed in the remanent flow  after agitation has stopped. In the case of highly viscous fluids, on the other hand, the fluid motion is promptly arrested by viscous dissipative forces and fluid mixing by a spoon is inefficient. Note, however, that mixing can still be achieved if the spoon were to perform an elaborated oscillating motion \cite{gouillart2006}. Without the proper selection of the impeller and characterization of the flow patterns it creates for a given  configuration, pumping and mixing at low Reynolds numbers can become very expensive in terms of time or power consumption. Furthermore, the resulting flow will often have  regions of poor fluid transport or even stagnant zones
\cite{Dickey2000,Bakker1995,Tanguy1997,Galindo1992}.

In Nature, there are many instances which involve flow without inertia. Many micron-size  organisms need to move and feed in an environment at low Reynolds number \cite{Purcell1977}.   Due to the time-reversibility of the Stokes equations governing the fluid motion at low Reynolds number, the only way a microorganism can produce net propulsion is by exerting a swimming stroke which is not identical under a time-reversal operation, i.e.~which does not follow the same sequence of strokes when the time runs forward or backward -- the so-called scallop theorem   \cite{Purcell1977}. The propagation of waves is the typical nonreciprocal motion and is the mechanism many microorganisms use to swim \cite{Taylor1951,Brennen1977,Childress1981,Lauga2009}. Traveling waves can be produced either by rotating a relatively stiff helical-shaped filament  or by passing bending waves along a flexible flagella in a time periodic fashion \cite{Brennen1977,Lauga2009}.

Swimming and pumping are two inter-related phenomena and in general a swimmer held in place will exert a net force on the fluid and pump \cite{Raz2007}. Inspired by  locomotion  due to  flexible flagella, in this study we investigate experimentally the pumping capacity of a  simple impeller consisting in a flexible elastic flapper that oscillates in a highly viscous Newtonian fluid. We start below by presenting a short literature  review of the different strategies that have been employed to pump or mix liquids at low Reynolds  numbers in order to put in perspective the particular strategy proposed in our paper.

\section{Background and structure of the paper}

\begin{figure}
\centering
\includegraphics[width=0.9\textwidth]{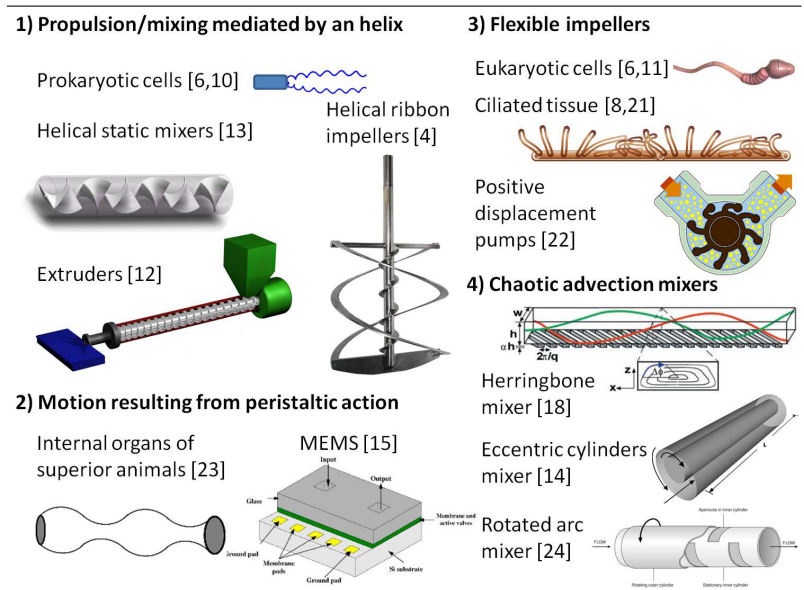}
\caption{Different strategies  to actively transport fluid at low Reynolds number. Four main classes have been recognized in the literature: transport assisted by an helix, peristaltic motion, flexible impellers (also found in positive displacement pumps), and chaotic-advection mixers. This classification is not unique and there are  other ways to transport fluid, including active and passive geometries,  in
microfluidic devices, see for example Ref.~\cite{Lee2011}. All images were reproduced with permission: helical ribbon impeller figure courtesy of Fusion Fluid Equipment, www.fusionfluid.com;
peristaltic micropump figure courtesy of Teymoori \& Abbaspour-Sani \cite{Teymoori2005}. Copyright 2005 Elsevier;
staggered herringbone mixer courtesy of Stroock \etal \cite{stroock202}. Reprinted with permission from AAAS;
eccentric cylinders mixer figure courtesy of Rodrigo \etal \cite{Rodrigo2003}. Copyright 2003, AIP Publishing LLC;
rotated arc mixer figure courtesy of Metcalfe \etal \cite{Metcalfe2006}. Copyright 2005 Wiley.} \label{Prop_Mix}
\end{figure}

Fluid transport at low Reynolds numbers has received considerable attention due to the demand for effective pumping and mixing operations assigned to process polymeric fluids, as is in the case of the food, pharmaceutical, or cosmetic industries \cite{Yerramilli2004,Thakur2003,Rodrigo2003}.  Recently, the development of microfluidic devices and microelectromechanical systems (MEMS) has brought renewed interest to  the field  \cite{Teymoori2005,Ottino2004,Bottausci2004,Brody1996}. We show in Fig.~\ref{Prop_Mix}  some examples of natural and man-made actuators that pump, mix or propel in a fluid at low Reynolds numbers. The use of rotating rigid helices is common. Prokaryotic cells use this type of actuation in order to swim \cite{Lauga2009}, and if bacteria are artificially tethered on a surface they are able to pump fluid \cite{Kim2008}. Helices are further used in extruders to transport viscous fluids \cite{Yerramilli2004}, and finally in mixers, either mobile \cite{Tanguy1997} or static \cite{Thakur2003}.  Flexible actuators are used by eukaryotic cells (such as  spermatozoa) to swim and also to pump biological fluids as is in the case of the collective motion of  cilia found in many superior animals \cite{Sleigh1988}. Flexible impellers have also been employed in the design of positive displacement pumps used in the industry \cite{Johnson}. The flexible impeller studied in this paper  represents a kind of ``elastic oar'' \cite{Purcell1977}.

Machin \cite{Machin1958} conducted a pioneering investigation to study the balance between elastic and viscous forces acting on flagellum. He modeled the physical balance between elastic forces and viscous drag proposing a elasto-hydrodynamic model \cite{Wiggins1998}. For weak viscous forces the filament is stiff and its motion is nearly reciprocal, producing no net motion. On the other hand, when the bending forces are weak, the viscous stresses prevent the formation of a bending wave on all of the flagellum but close to the actuation region, and therefore, an optimum stiffness/drag balance is expected to occur when both effects are of comparable magnitude.  For a periodic actuation at frequency $\omega$, this balance is quantified by an elastohydrodynamic penetration length, $\ell$, defined as  \cite{Wiggins1998,Gauger2009}
\begin{equation}\label{elastohydro}
\ell={\left(\frac{EI}{\omega\zeta_{\bot}}\right)}^{{1}/{4}},
\end{equation}
where $EI$ is the filament bending stiffness and $\zeta_{\bot}$ is the viscous drag coefficient that acts normal to the local velocity on the flagellum. The length $\ell$ should then be compared with the total length of the filament, $L$. A dimensionless number,  $L/\ell$, compare the two -- it is   colloquially known as the Sperm number, $\rm Sp$ \cite{Lowe2003}. In several experimental and computational studies, a maximum swimming speed, or pumping performance, has been obtained  for $1\lesssim{\rm Sp}\lesssim4$ \cite{Wiggins1998,Gauger2009,Lowe2003,Dreyfus2005,yu06,lauga07_pre,Onshun2011,espinoza2013}.

The idea of our paper is to use this concept of optimal balance between viscous forces and bending resistance to  produce non-reciprocal deformation of a flexible flapper and thus pumping  in  creeping flow.  The  paper is organized as follows. We first describe in section \ref{exp} the fluid used in the investigation and  describe  the flapper actuation and properties. We then present the flow measurements and the quantification of pumping in section \ref{measure}. Our  experimental results are discussed in  section \ref{results}. In this section, an analysis and discussion of the pumping capacity  for different values of $L/\ell$ is also presented. The conclusions are presented in section \ref{conclusion}.

\section{Experimental setup}
\label{exp}

\begin{figure}
\centering
\includegraphics[width=0.9\textwidth]{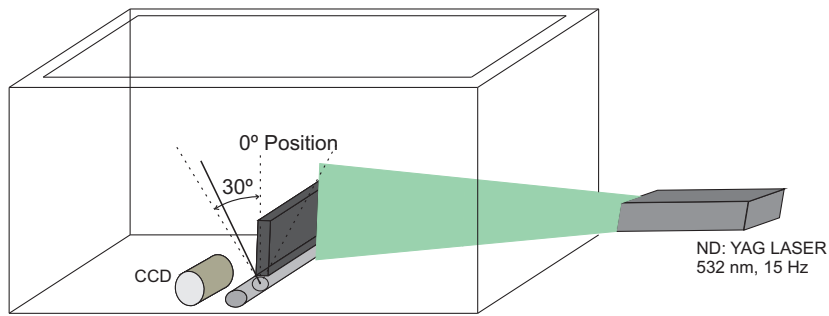}
\caption{Schematic representation of the flapper and the Particle Image Velocimetry setup.}
\label{cajaPIV}
\end{figure}

The experiments were conducted in a rectangular tank, shown schematically in Fig.~\ref{cajaPIV}, filled with a highly viscous fluid. A rectangular flapper was placed in the middle of the tank, attached to shaft  driven by a step motor. The tank dimensions (length 35 cm,  height 24 cm, and width 15 cm) and the arrangement of the flapper allowed to produce a nearly two-dimensional flow in the measurement zone (middle of the container). A lid was placed to keep the tank closed and prevent glucose crystallization and deformation of the fluid surface due to the flapper motion \cite{Trouilloud}. Flappers with different levels of flexibility were fabricated by cutting an elastic  sheet into a rectangular shape with fixed width, ($w$ =10 cm). The length of the flappers was $L$= 5 cm in most cases and a few tests with shorter, stiffer flappers were conducted to validate the scallop theorem.  The properties of all flappers used in our experiments are shown in Table \ref{flappersmaterials}. The flexibility was varied by selecting materials with different elastic rigidity and also by varying the thickness, $t$, of the sheet.

A highly viscous Newtonian fluid was prepared  by adding 6\% of water to glucose (${}^{\circ}$B\'e=45, 3000 Pa$\cdot$s at 23${}^{\circ}$C).
The fluid was mixed for 48 hours at very low speeds and let to rest for 3 days to ensure homogenization. The liquid viscosity was measured to be  63 Pa$\cdot$s  with an Anton Paar \emph{Physica MCR 101} rheometer, and a density of 1540 kg/m$^3$  was measured with a pycnometer (25 ml). The  high viscosity ensured  that all  experiments remained in the low Reynolds number regime. To ensure  that the liquid was Newtonian,  oscillatory rheological tests were conducted (not shown) and for the range of shear rates in which the pumping experiments were conducted, the storage modulus was negligible compared with the loss modulus.

\begin{figure}[t]
\centering
{\mbox{\subfigure[]{\includegraphics[width=0.38\textwidth]{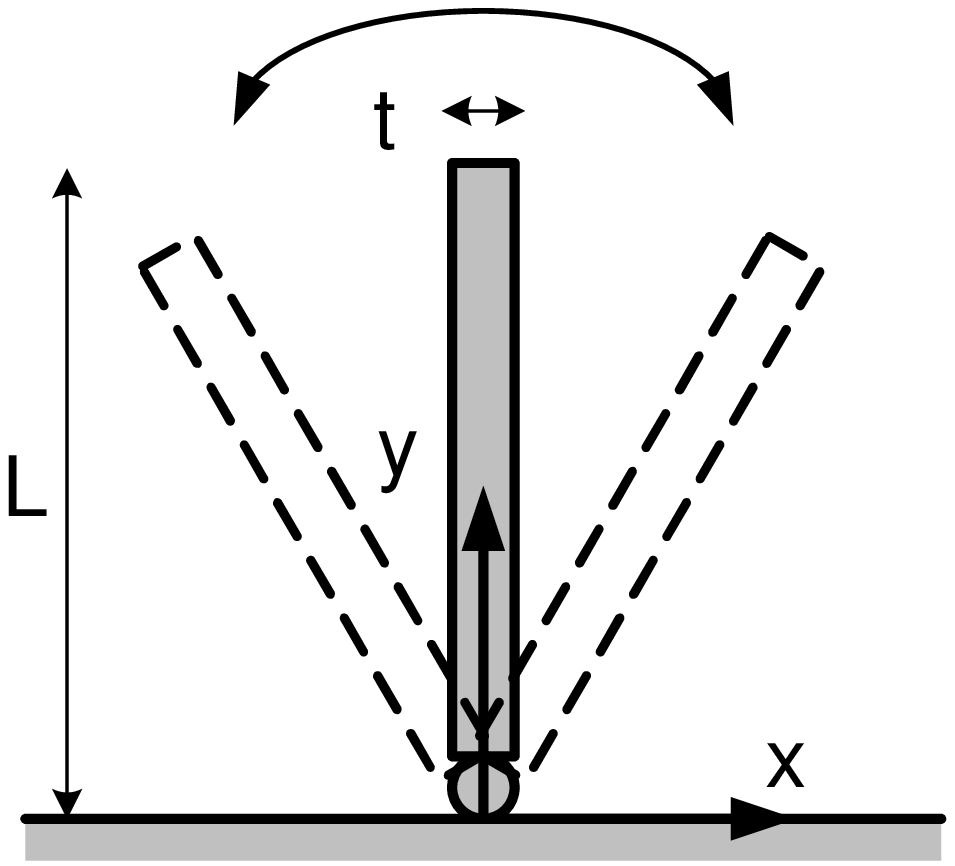}}}}\quad\quad
{\mbox{\subfigure[]{\includegraphics[width=0.34\textwidth]{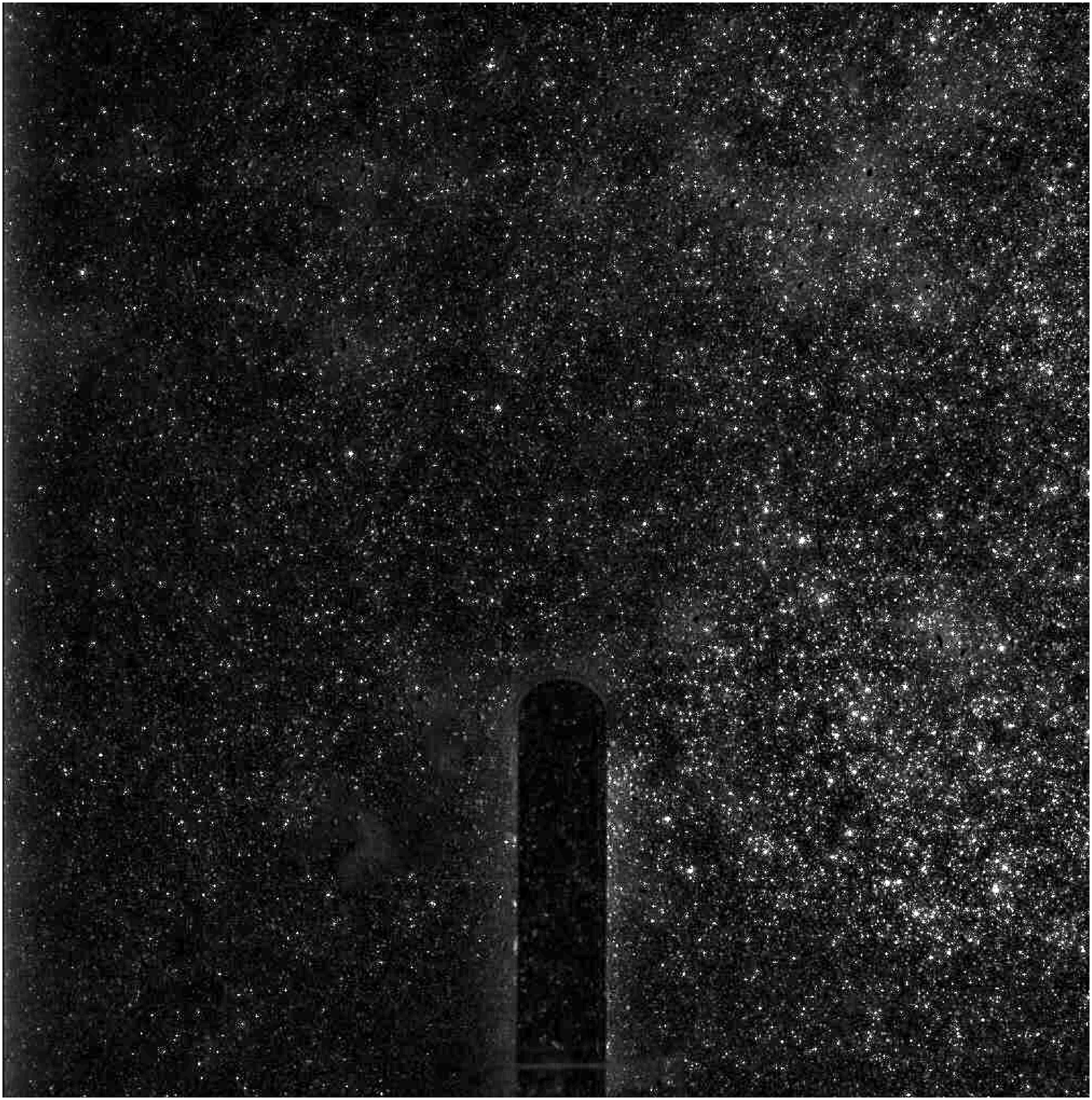}}}}
{\mbox{\subfigure[]{\includegraphics[width=0.42\textwidth]{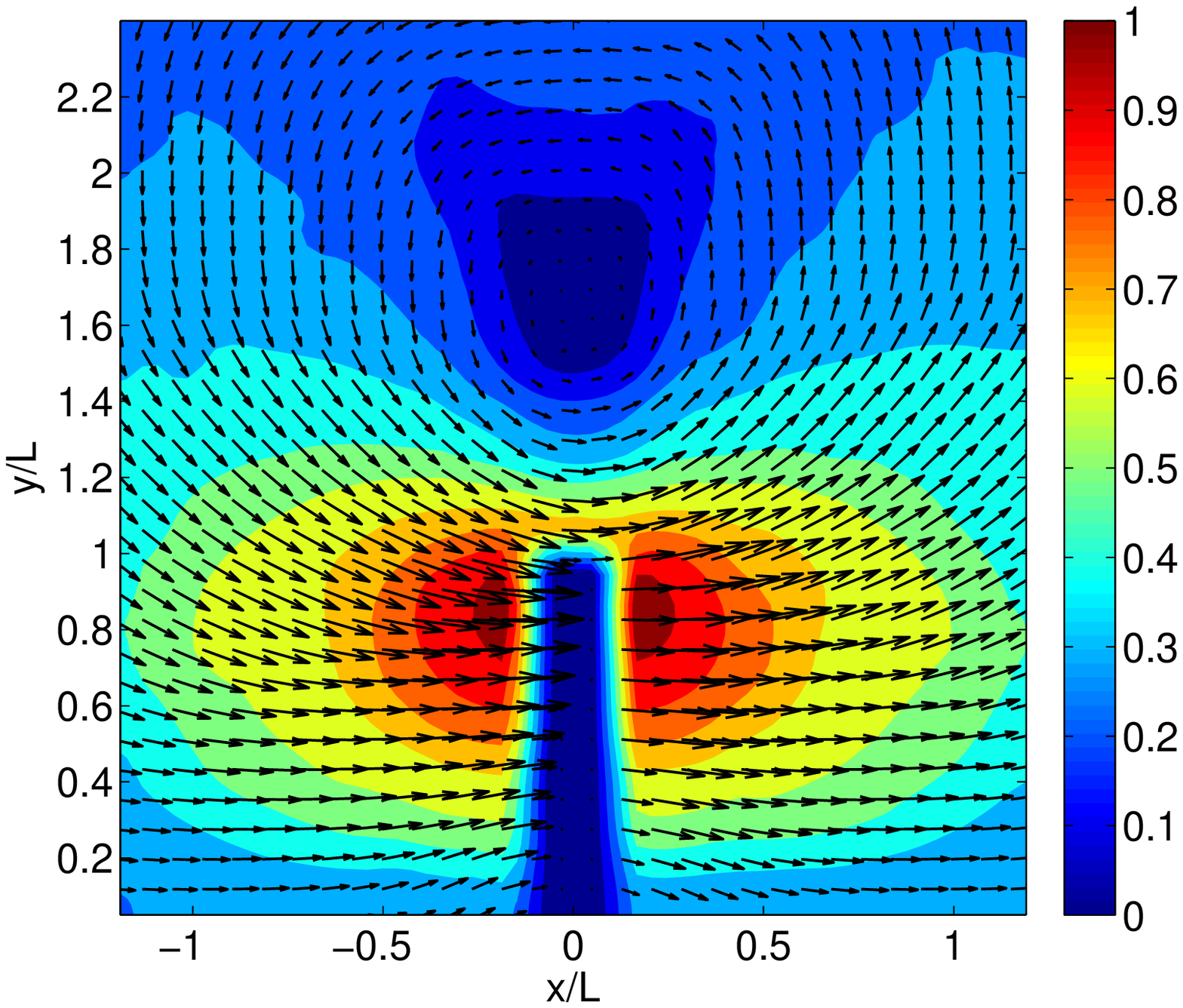}}}}
\caption{(a) Schematic description of one oscillation period of the flapper; (b) Typical snapshot obtained from the PIV system; (c)  Typical instantaneous velocity field and iso-velocity contour (normalized by the velocity of the flapper tip).}
\label{fig:flapper_motion}
\end{figure}

The oscillatory motion of the flapper was generated using a stepper motor (\emph{UE73PP}) of 4 phases, placed outside the tank and connected to the rotation axis of the flapper through a wall connector. The oscillatory amplitude was fixed at $\pm$30 degrees from the vertical. The angular velocity was varied using a motion controller (\emph{Newport MM4006}) operated with the  software \emph{LabView 8.1}. The motion of the flapper is depicted schematically in Fig.~\ref{fig:flapper_motion}(a).

\begin{table}[t]
  \centering
  \begin{tabular}{|c|c|c|c|c|}\hline
   Flapper &  $t$, mm & $L$, mm & $E$, MPa &  $EI$, Pa m$^4$ \\\hline
   aluminum (Al)&6.1 & 28.5 & 7000.0 & 132.42  \\
   acrylic (Ac), $\bullet$ & 8.5 & 50 & 2240.0 & 11.4637 \\
   neoprene (N1),{\color{blue} $\blacksquare$} & 6.4 & 50 & 2.73 & 5.96 $\times 10^{-3}$ \\
   neoprene (N2), {\color{red} $\circ$} & 3.2 & 50 & 2.73 & 7.45 $\times 10^{-4}$\\
   neoprene (N3), {\color{red} $\ast$}& 1.2 & 50 & 2.73 & 3.93 $\times 10^{-5}$\\
   silicon rubber (SR), {\color{red}$\square$} & 0.8 & 50 & 2.15 & 9.17 $\times 10^{-6}$\\\hline
  \end{tabular}
  \caption{Physical properties of the flappers used in this study. For all flappers, the width is $w = 100$ mm, $t$ is the flapper thickness, $L$ its length and $E$ is the Young modulus. Denoting $I$  the second moment of inertia, $I=wt^3/12$, we have $EI$ as the bending rigidity. }
  \label{flappersmaterials}
\end{table}

For each flapper,  three different velocities were tested. The velocity of the flapper is characterized by $V_{\rm tip}= \omega L$, where $\omega$ is the angular velocity of the flapper and $L$ is its length. Considering these speeds and fluid properties, the Reynolds number, Re=$V_{\rm tip} L \rho/\mu$, was smaller than 0.05 in all cases. Lowering the Reynolds number even further was not possible without increasing the power of the step motor used in the arrangement.

A standard particle image velocimetry (PIV) technique was used to obtain the velocity field generated in the liquid, also shown schematically in Fig.~\ref{cajaPIV}.  A Nd:YAG laser system generates a 50 mJ energy 532nm laser beam which was converted to a laser sheet using optics. The laser sheet illuminated a vertical slice at the center of the tank. A CCD camera was positioned to record images illuminated by the laser sheet. The resolution of the camera was $1008\times1016$ pixels and the typical measurement area was $121\times122$ mm$^2$. Silver-coated glass spheres with an average diameter $10\pm5 \mu m$ were used as particle tracers. A typical image of the seeded flow is shown in Fig.~\ref{fig:flapper_motion}(b).  The velocity field consisted of $62\times62$ vectors using an interrogation area of $32\times32$ pixels and an overlap of $50\%$. The spatial resolution was $2.24\times2.24$ mm$^2$ for most of the experiments. For all cases, considering our setup, the displacement never exceeded 7 pixels/cycle. Standard filtering and validation of the velocity fields were applied to all measurements.

Clearly, due to the effect of containing walls and finite size of the flapper, some slight deviations from a pure two-dimensional flow could be expected. To evaluate how close our measurements are from being two-dimensional we can calculate the two-dimensional divergence of the flow everywhere in the measurement area: $D=\partial u/\partial x + \partial v/\partial y$. Small values of $D$ would indicate that the gradient of the velocity component perpendicular to the measuring plane are small. To estimate the order of magnitude of the velocity in that direction we can calculate $DW_{laser}$, where $W_{laser}(\approx 0.8$ mm) is the thickness of the laser plane. For the typical measurements show in Fig. \ref{fig:velocity_fields}, the average value of $DW_{laser}/V_{tip}$ is $1.7\times10^{-4}$ in the region where the pumping occurs. The measurements presented here are therefore very close to being two-dimensional.

\section{Flow and pumping measurements}
\label{measure}
The fluid motion  resulting from the oscillation of the flapper was quantified in two different ways. The first one inferred the flow velocity field from images obtained from the seeded fluid in closely spaced time instants. This is the typical PIV measurement and it allowed to obtain  the instantaneous velocity field around the flapper. An example of such measurement is shown in Fig.~\ref{fig:flapper_motion}(c), for the particular instant in which the flapper is in its vertical position and is moving in the clockwise direction.

\begin{figure}[t]
\centering
{\mbox{\subfigure[Rigid flapper]{\includegraphics[width=0.45\textwidth]{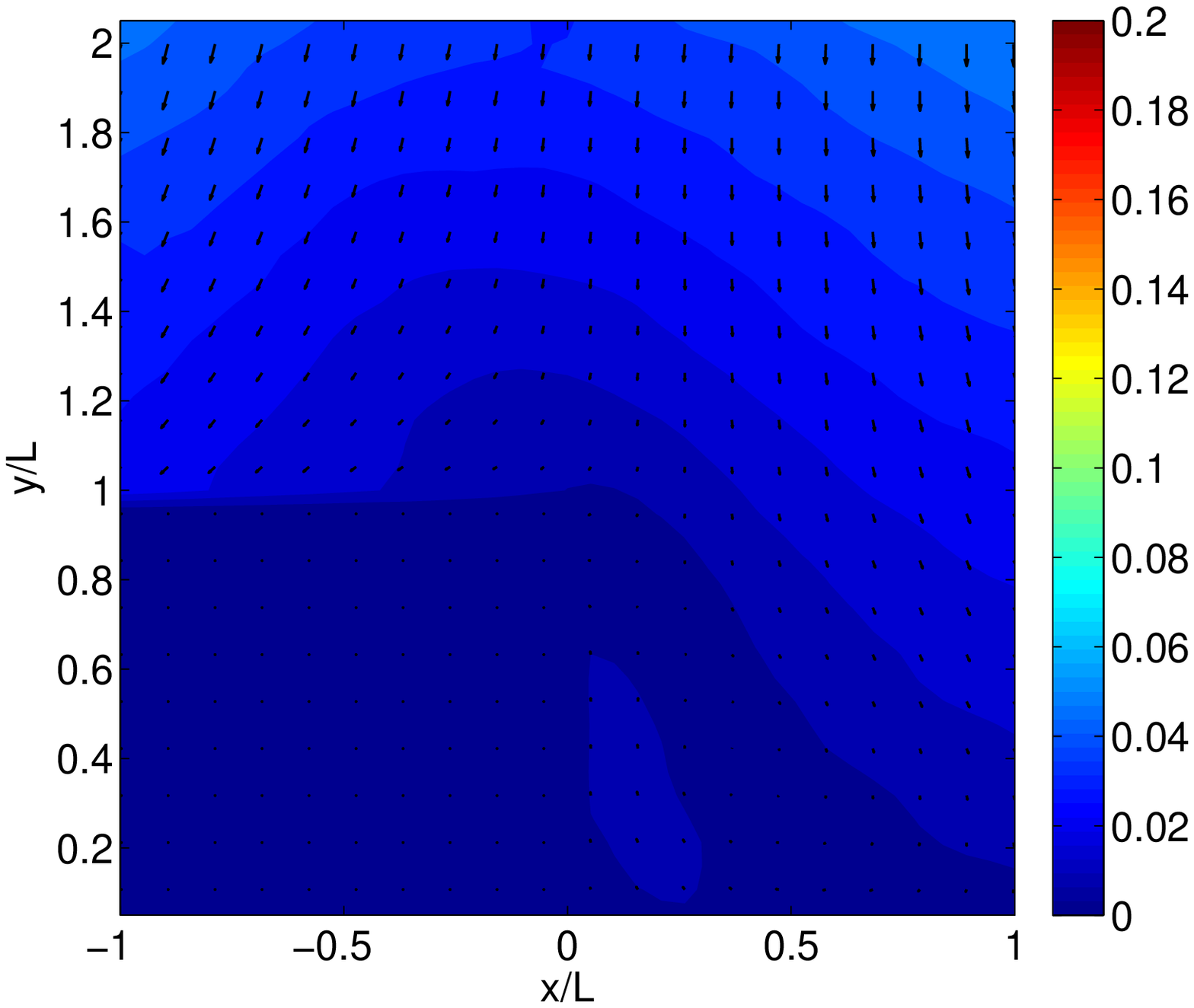}}}}\quad\quad
{\mbox{\subfigure[Flexible flapper]{\includegraphics[width=0.45\textwidth]{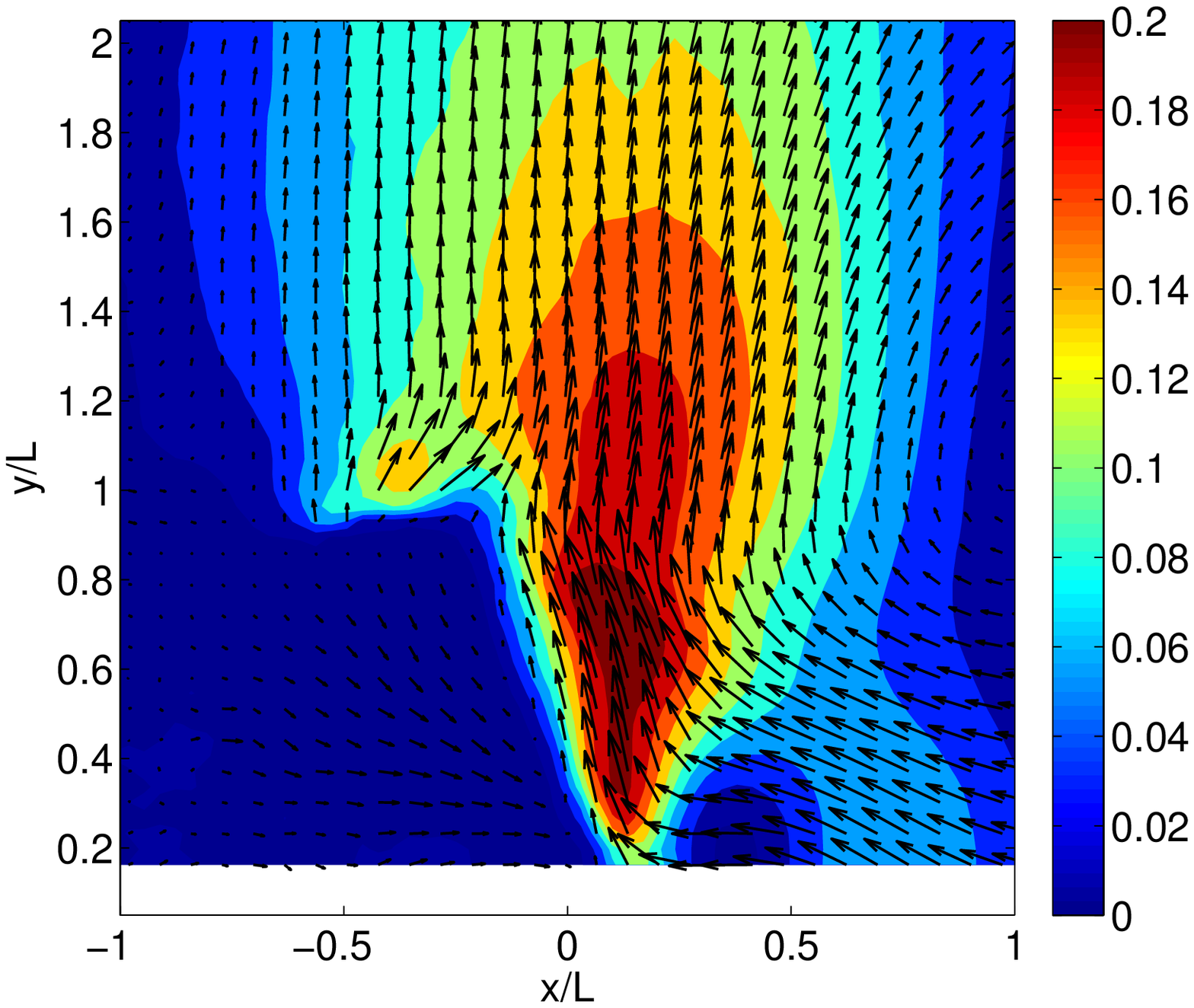}}}}
\caption{Mean velocity fields produced after one flapping period. The colors indicate the magnitude of the vertical velocity, normalized by $V_{\rm tip}$. The flexible and rigid flappers correspond to cases (N2) and (Al), as described in Table \ref{flappersmaterials}.}
\label{fig:velocity_fields}
\end{figure}

However, this particular determination of the flow field does not provide a measure of the net fluid motion produced in each cycle, since most of the fluid transported during one half of the cycle is expected to be transported the other way during the second half. Specifically, if the flapper is rigid, and the flow is for ${\rm Re}\ll1$, the flow field at the instant where the flapper is vertical and moves  in the counter-clockwise will be the exact   mirror image of the one shown in Fig.~\ref{fig:flapper_motion}(c). Therefore, in order to determine the net fluid motion induced by the flapper over an entire oscillation cycle, a different strategy has to be used.

To address this issue, we took images of the tracer particles  only when the flapper hinge passed through the vertical position in the clockwise direction. In other words, an image was obtained only once during the entire oscillating cycle, separated by the oscillation period (denoted $T$). By applying the cross-correlation algorithm to pairs of images obtained a time $T$ apart, the net fluid motion per oscillation cycle was obtained. An example of the measured velocity field resulting from this technique is shown in  Fig.~\ref{fig:velocity_fields}. For all experiments conducted in this investigation, 100 realizations (oscillation periods) were considered in order to obtain statistically converged measurements of the velocity field. The uncertainty of the measurement was determined from the standard deviation of the data set for each condition.

\begin{figure}[t]
\centering
{\mbox{\subfigure[Rigid flapper]{\includegraphics[width=0.45\textwidth]{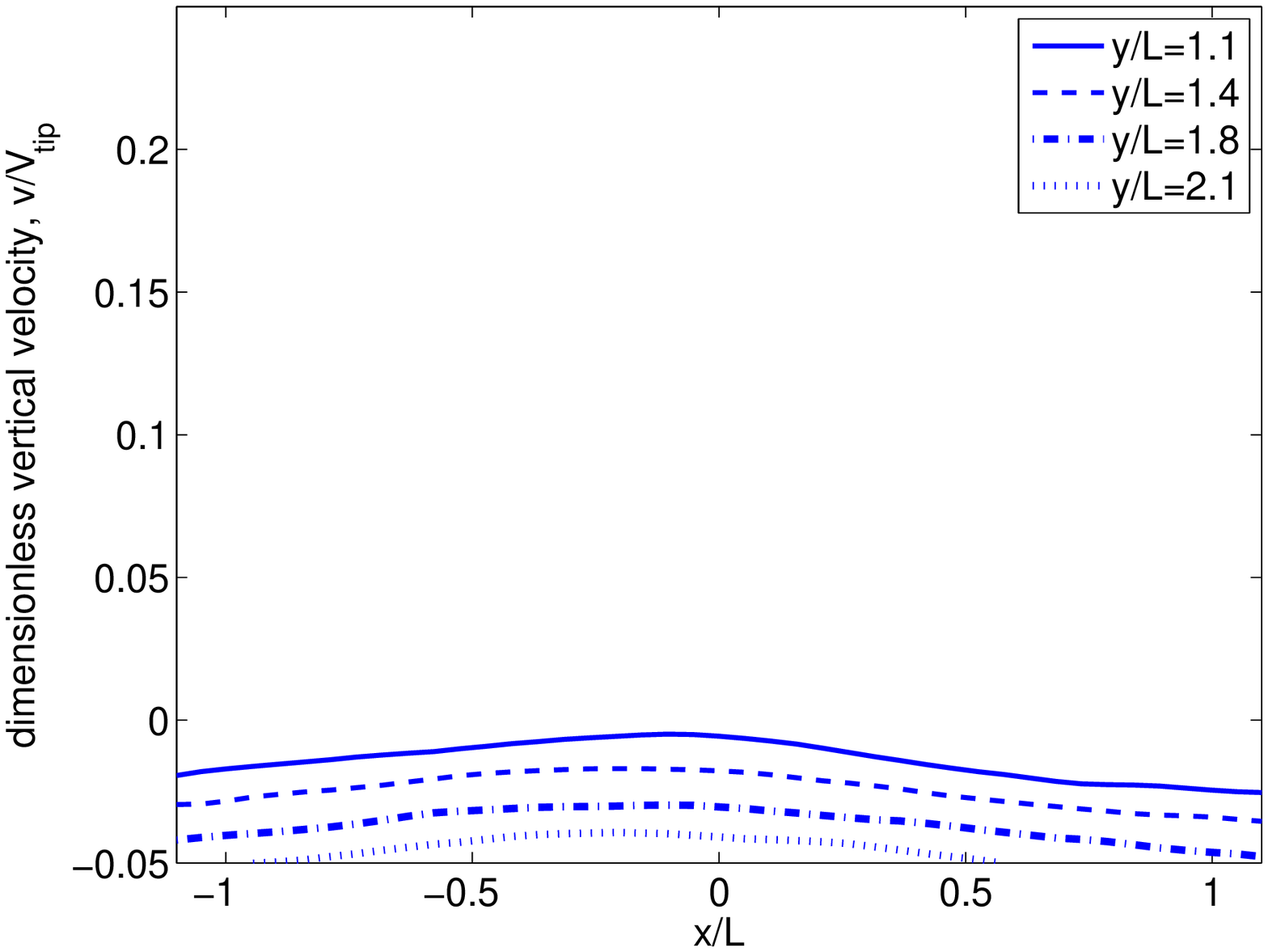}}}}\quad\quad\quad
{\mbox{\subfigure[Flexible flapper]{\includegraphics[width=0.45\textwidth]{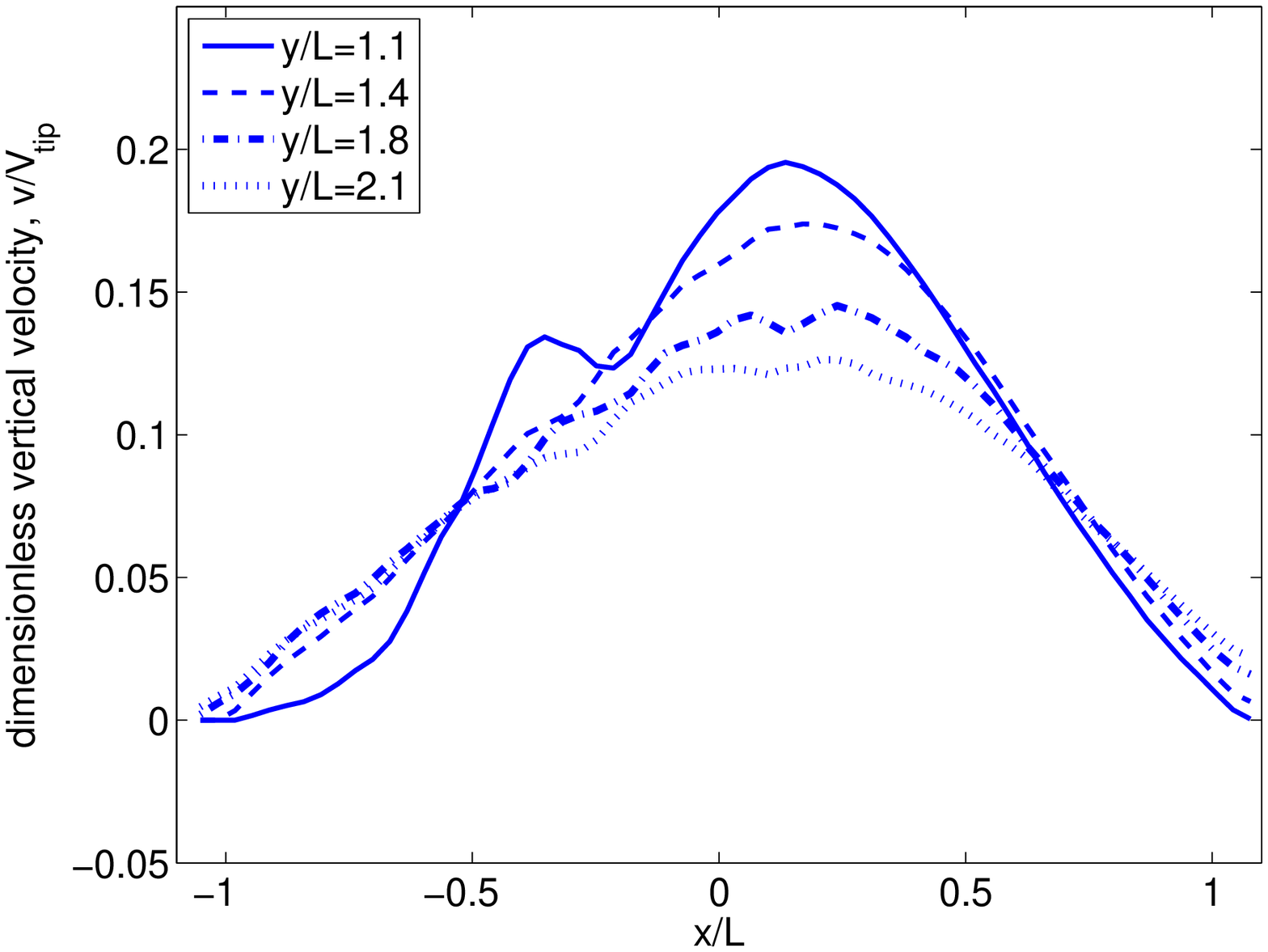}}}}\\
{\mbox{\subfigure[Pumping]{\includegraphics[width=0.45\textwidth]{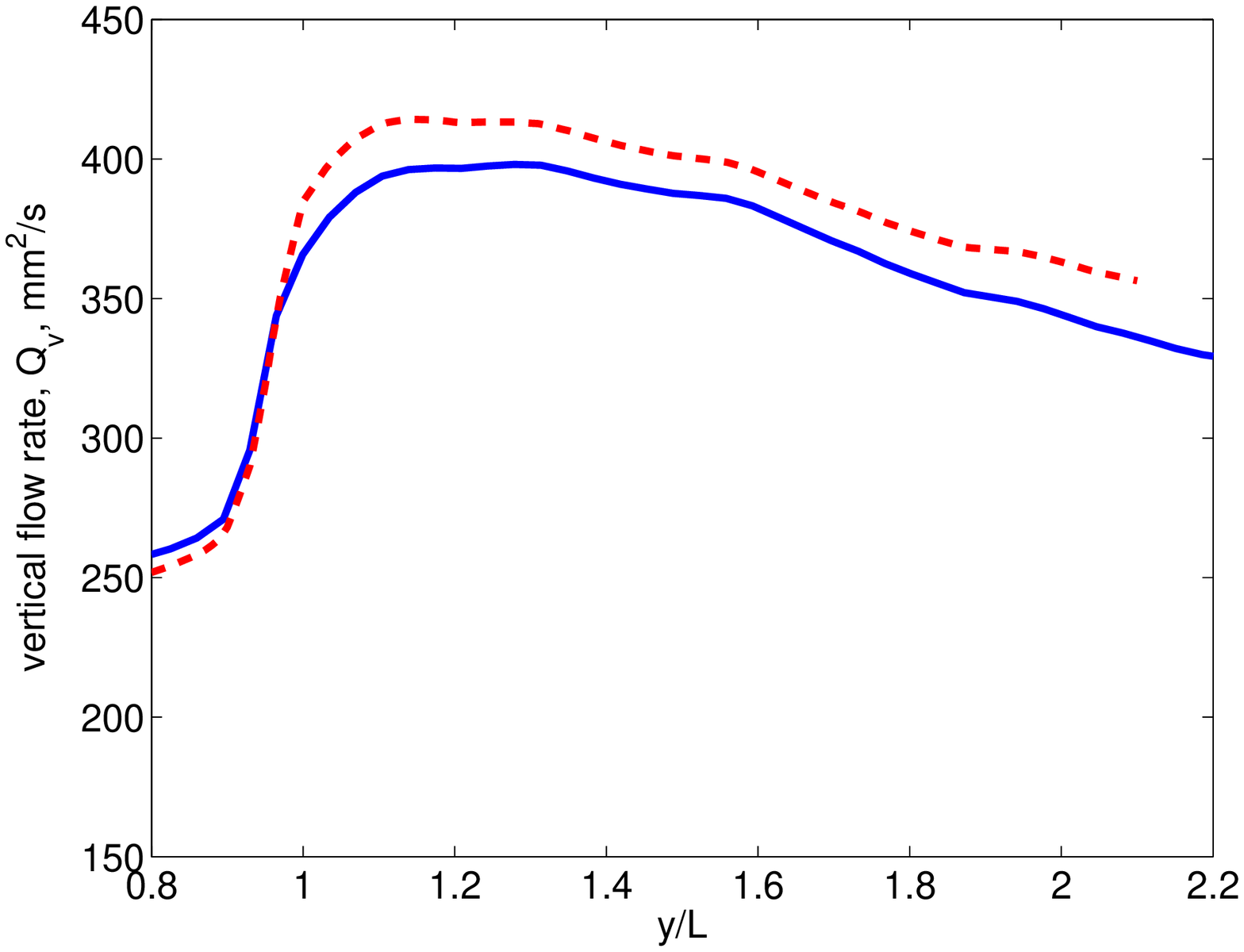}}}}
\caption{(a-b): Typical vertical velocity profiles for different vertical distances above the flapper tip. These profiles correspond to the fields shown in Fig.~\ref{fig:velocity_fields} for rigid (Al) and flexible (N2) flappers (a and b, respectively). (c) Measured net flow rate, $Q_v$,  as a function of dimensionless vertical distance, $y/L$, as calculated by integrating the data from (b). The dashed line shows the value of $Q_v$ when the effect of sedimentation is considered.}
\label{fig:velocity_profiles}
\end{figure}

In order to  quantify the amount of liquid set in motion by the flapping impeller, the net flow rate was calculated. Since the time-averaged velocity field is known, we define the pumping capacity as the integral
\begin{equation}\label{pumping}
Q_{v}=\int_{x_{1}}^{x_{2}}v\,dx,
\end{equation}
where $v$ is the vertical velocity component of the cycle-averaged velocity field and $x$ is the horizontal coordinate (along the bottom of the tank and perpendicular to the average flapper orientation). The quantity $Q_{v}$ measures the flow rate in the vertical direction, and its value depends on the values of the bounds of the integral, $x_{1}$ and $x_{2}$, and also on the vertical position, $y$, at which it is evaluated. If the measurement was conducted along the entire horizontal length of the container ($x_2-x_1= L_c$, where $L_c$ is the container length), the result would be $Q_v=0$ since the upward current induced by the flapper would be balanced  by the returning fluid going downwards. In order to  account solely  for the net flapper-induced current, we chose $x_1$ and $x_2$ to coincide with the extremities of the region where the vertical velocity has a positive value, i.e.~we only account for the upward pumping. We show in Fig.~\ref{fig:velocity_profiles}a \& b   typical velocity profiles for the vertical velocity as a function of the horizontal coordinate, at different  distances, $y$,  above  the tip of the flapper. The flexible flapper (Fig.~\ref{fig:velocity_profiles}b) generates a vertical upward jet of a flow rate measured thus by $Q_{v}$.
The region where the measurement is taken  extends from $x_1\approx-L$ to $x_2\approx L$ for all cases (recall that $L$ is the flapper length). In Fig.~\ref{fig:velocity_profiles}c we then plot the value of $Q_v$ measured at different vertical distances, $y$, above the  flapper tip. The net flow rate in the upward jet   reaches a maximum value right above the flapper and it then slowly decreases away from the jet. That decrease results from the increase in the  width of the induced jet beyond the measurement region. We use, as a measure of the  pumping capacity for each flapper, the maximum vertical value for $Q_v$.

It is important to note that our measurements are affected by the sedimentation of the tracers. Since sedimentation acts in a direction opposite to that achieved by flexible flapping, the measurements are slightly smaller than what they should be. To estimate the effect of sedimentation, we can simply add the two vector fields (Fig.~\ref{fig:velocity_fields} a and b). We must note that this procedure is not strictly correct since the velocity fields are time averages; also, the position of the flapper in both cases is not exactly the same. Nevertheless, by adding the fields we can estimate an order of magnitude of the error that results from neglecting the effect of tracer sedimentation. The dashed line in  Fig. \ref{fig:velocity_profiles} (c) shows the flow rate calculated from the `corrected' velocity field (considering the same experimental conditions as the continuous line). The trend of the measurement is not affected significantly but the magnitude is slightly increased. By adding the effect of sedimentation the flow rate in the region where the measurement is taken ($y/L\approx 1.2$) increases by about 4.3\%. Therefore, we can expect our  pumping measurements to be underestimated by a few percent.

\section{Results and discussion}
\label{results}
The goal of our paper is to demonstrate that when  the flapper is made of a flexible material, a net flow is induced by the periodic motion. Below we first discuss  the nature of the induced flow, and then  show how the pumping capacity changes with the flapper tip speed and rigidity.

\subsection{Rigid vs.~flexible}
We display in Fig.~\ref{fig:velocity_profiles}a \& b the typical profiles of the net fluid velocity field during an oscillation cycle for two distinctively different flappers, rigid (a) and flexible (b).  In the rigid case we observe no upward flow but a small downward flow resulting from   the slow sedimentation of the tracer particles used for the PIV measurements. We conducted an additional experiment  in which we left the flapper fixed and obtained PIV measurements of the flow field, and obtained  results identical to  that in
Fig.~\ref{fig:velocity_profiles}a. Our rigid results validate therefore experimentally  the scallop theorem: the reciprocating motion of a rigid flapper in creeping flow should not induce any pumping.

The vertical flow field induced from a typical flexible flapper is shown in Fig.~\ref{fig:velocity_profiles}b. Clearly, in this case, there is a significant amount of fluid moving upwards on average. Notice that in the figure, the velocity field is slightly slanted to the left. Our flow measurements are computed  from images obtained when the flapper is in its vertical position and moving in the clockwise direction. However, since the flapper is flexible, when the base of the flapper is indeed positioned at zero degrees, its tip is lagging behind and curved to the left.  That delay of the tip with respect to the base is in fact what induces the net vertical fluid motion, and it is responsible for the slight asymmetry observed.

\subsection{Understanding flexible pumping}

\begin{figure}
\begin{center}
{\mbox{\subfigure[Rigid flapper (Ac)]{\includegraphics[width=0.45\textwidth]{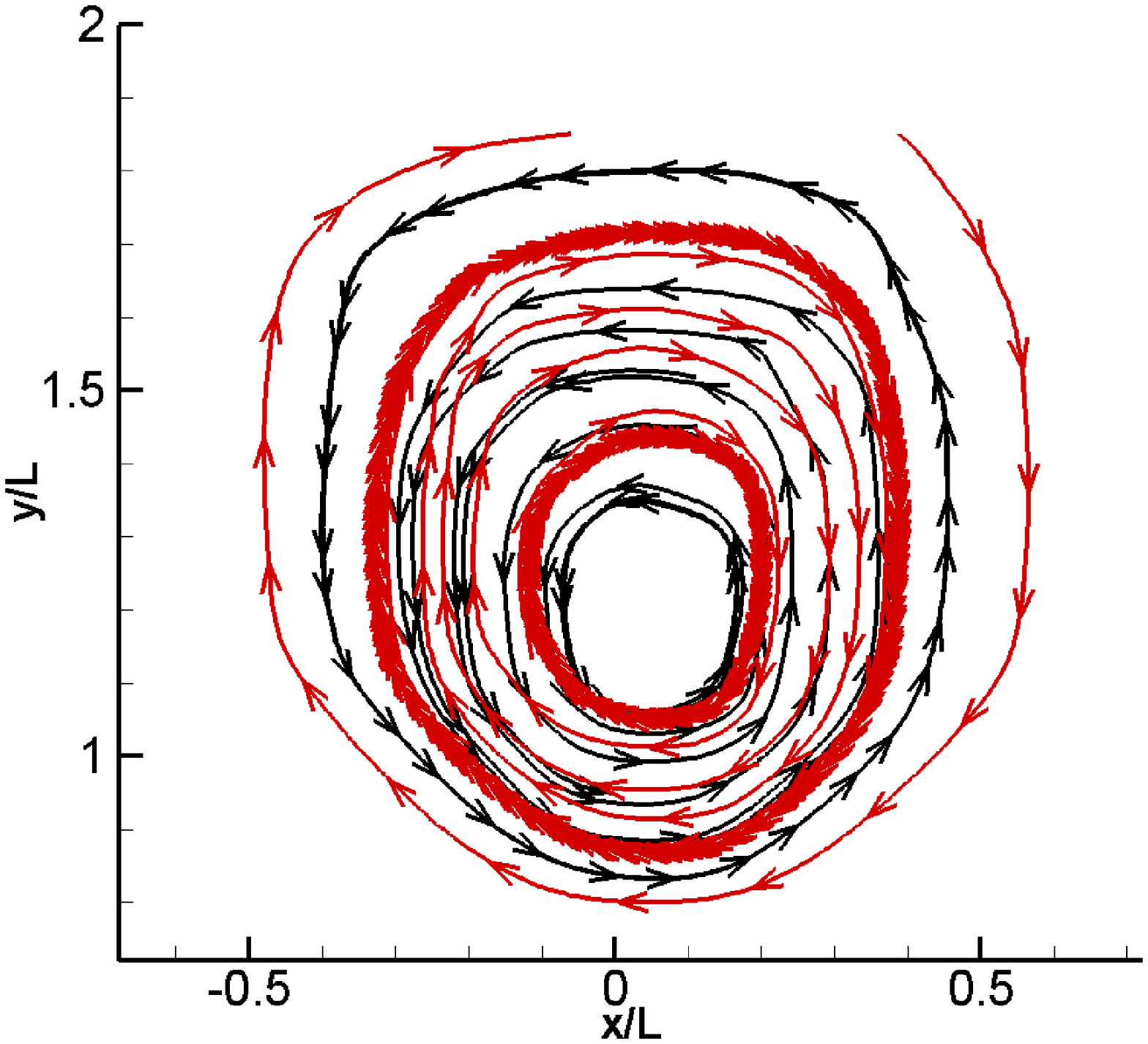}}}}\quad
{\mbox{\subfigure[Flexible flapper, (N2)]{\includegraphics[width=0.45\textwidth]{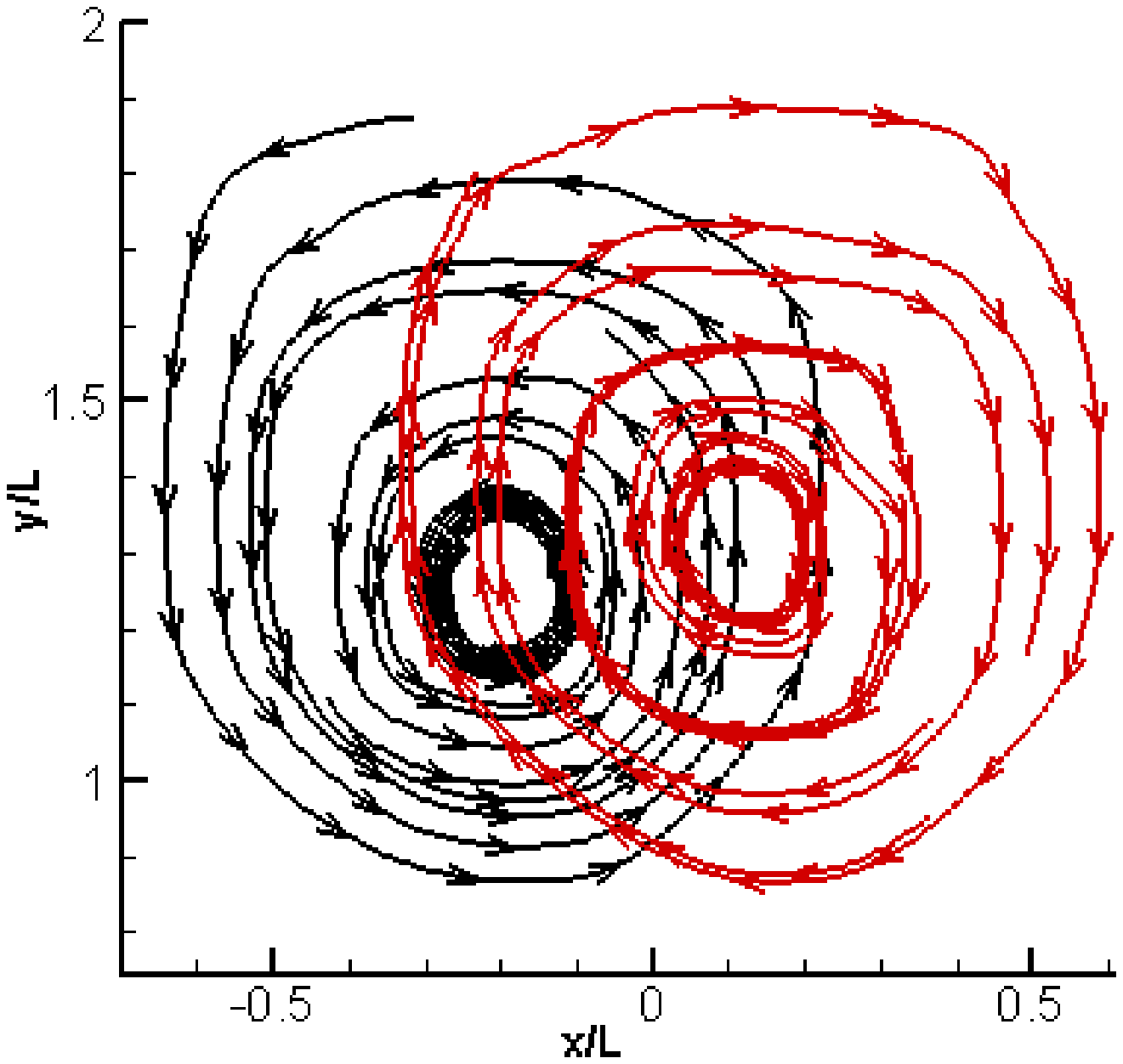}}}}\\
\end{center}
\caption{Instantaneous streamlines formed in the fluid with the rigid flapper (Ac in Table\ref{flappersmaterials}) and flexible flapper (N2 in Table\ref{flappersmaterials}). In both cases the tip speed is $V_{\rm tip}$ = 5.48 mm/s.  Black lines: streamlines formed during the clockwise rotation; red lines: streamlines formed during the counterclockwise rotation.} \label{breaksymmetry}
\end{figure}

In order to further understand the process that produces the vertical flux, we analyze the instantaneous velocity fields  shown in  Fig.~\ref{fig:flapper_motion}c for two specific  flappers. In Fig.~\ref{breaksymmetry} we plot the  instantaneous streamlines formed by the rigid flapper (case Ac in Table \ref{flappersmaterials}, streamlines in  Fig.~\ref{breaksymmetry}a) and the  flexible N2 flapper (Fig.~\ref{breaksymmetry}b)  at the moment where the  base of both flappers is oriented vertically, either in the clockwise (black lines) or counterclockwise (red lines) directions.

A large vortical structure can be observed above the tip of the flapper.
In the case of the rigid flapper (Fig.~\ref{breaksymmetry}a), the vortices centers are located nearly at the same position for both  rotation directions and the corresponding streamlines essentially overlap each other. On average, the flows cancel each other resulting in zero net transport of the fluid over a period of  oscillation.

On the other hand, in the case of a flexible flapper, a strong asymmetry in the two vortices is apparent. This is because the tip of the flapper lags behind the actuation point in every oscillation. As a consequence, the vortical structure induced when the flapper is moving in the clockwise direction is  shifted to the left. Conversely, when the flapper is moving in the counter-clockwise direction, the center of the vortex is displaced  to the right. These two velocity fields do not cancel each other out over a single oscillation period, which results in a net vertical motion. In the region immediately above the flapper tip, the streamlines of the velocity fields for both directions of  oscillation have the same direction and point vertically away from the flapper, explaining the direction of the jet. Note that there is a slight asymmetry in the position of the two vortices, which most likely results from the back-slash of the motor. Despite this asymmetry, the effect is very clear.

\subsection{Parametric dependence on frequency and flapper properties}

With the goal of devising the optimal pump, we now study how  the strength of this jet depends on the flapping frequency and the properties of the flapper.  As the flapper rigidity decreases, it should be expected that  the pumping capacity will increase as a consequence of an increased asymmetry between the two vortices in Fig.~\ref{breaksymmetry}. The difference will also be influenced by the oscillating speed of the flapper. However, if the flapper becomes too flexible, it will no longer be able to  overcome the viscous stresses on its surface and, will thus not be able to instantaneously move the fluid in the first place. There must therefore  be an optimal combination of flapper rigidity and speed to produce the maximum amount of pumping.

\begin{figure}
\centering
{\mbox{\subfigure[]{\includegraphics[width=0.45\textwidth]{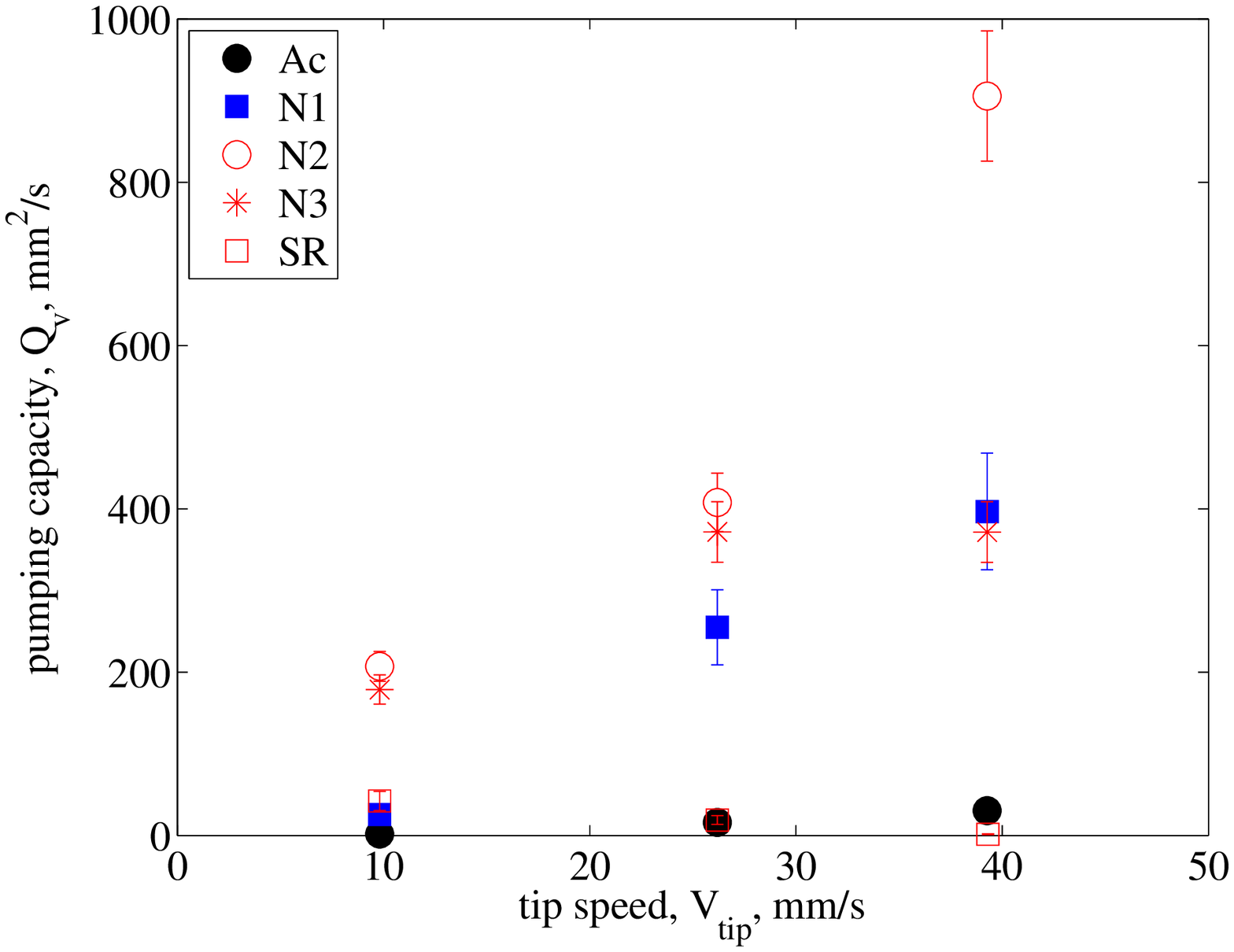}}}}\quad\quad\quad
{\mbox{\subfigure[]{\includegraphics[width=0.44\textwidth]{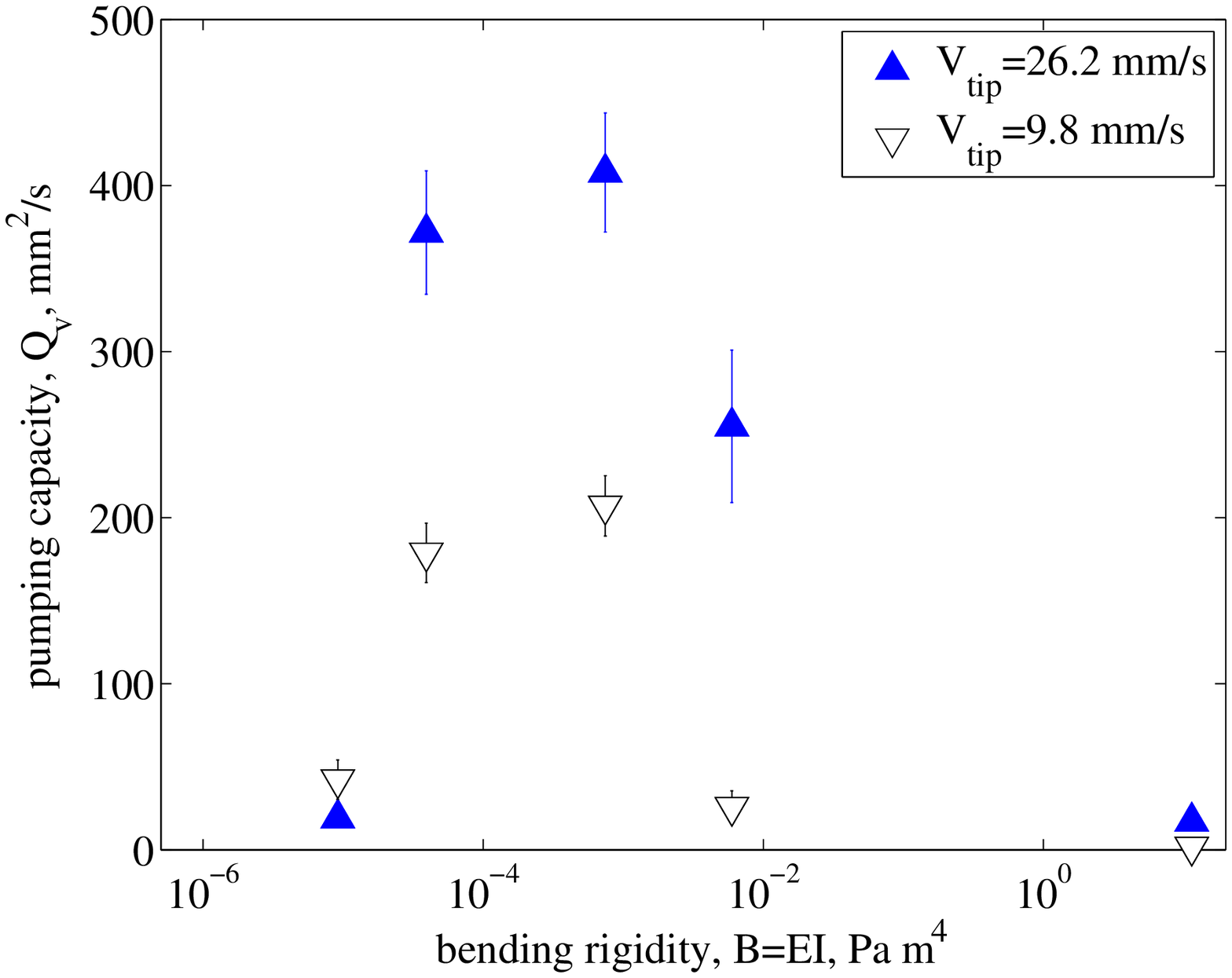}}}}
\caption{Pumping capacity as a function of the flapper tip speed (a) and  flapper rigidity (b). In all cases, the properties of the flappers are shown in Table \ref{flappersmaterials}. (a) Five different flappers: Ac, ($\bullet$); N1,({\color{blue} $\blacksquare$}); N2, ({\color{red} $\circ$}); N3, ({\color{red}$\ast$}); SR, ({\color{red}$\square$});
(b) Only one speed for each flapper (for each data group):  26.18  ({\color{blue} $\blacktriangle$}) and 9.82 mm/s ($\triangledown$).}
\label{fig:pump}
\end{figure}

In Fig.~\ref{fig:pump}a we plot the  pumping capacity, $Q_v$,  as a function of the tip flapper speed, $V_{\rm tip}$, for the five flappers tested in this study.
For the flapper with the highest rigidity (Ac,  filled black  circles), the induced pumping is essentially zero. As the rigidity of the flapper is reduced (N1, filled blue squares,  and N2, empty red circles), we observe that the pumping capacity increases monotonically with flapper tip speed, in the range of velocities tested. For even smaller rigidity (N3, red asterisks), the pumping capacity first increases with tip speed but decreases at high speed.

We then plot in Fig.~\ref{fig:pump}b the pumping capacity as a function of flapper rigidity, keeping the tip speed constant.
Data for a tip speed of  26.18 mm/s are shown as filled blue triangles while those with  9.82 mm/s are in empty down-triangles. For both data sets, we see that the pumping capacity is small in both limits of very flexible or very rigid flapper, and a maximum is reached at an intermediate value.

\subsection{Optimal flexible pumping}

Having demonstrated that it is possible to break the scallop theorem by using the  flapper flexibility, we  now recast our experimental results in a dimensionless manner in order to optimize pumping.

The pumping flow rate of the flapper can first be made dimensionless by using the flapper size and its tip velocity as
\begin{equation}\label{pumping2}
Q_{v}^{*}=\frac{Q_v}{V_{\rm tip}L},
\end{equation}
where $Q_{v}^{*}$ is now the dimensionless pumping capacity.
In order to describe the physical interplay between bending resistance and viscous forces, we also define a dimensionless number  using the viscous penetration length from  Eq.~\ref{elastohydro}. This dimensionless group, denoted $F$ here and termed a flapping number, physically measures the ratio of the flapper size, $L$, to $\ell$, and is thus defined as
\begin{equation}\label{flapping}
F\equiv L{\left(\frac{V_{\rm tip}\mu}{L EI}\right)}^{1/4},
\end{equation}
where $\mu$ is the shear viscosity of the fluid. This  flapping number is essentially the sperm number defined in the introduction and first proposed  by Machin \cite{Machin1958}, and we use a  different denomination here because our flapper is not a slender body. It is interesting to re-write this dimensionless group as
\begin{equation}\label{flapping2}
F=\left(\frac{L}{I^{1/4}}\right){\left(\frac{V_{\rm tip}\mu}{L E}\right)}^{1/4},
\end{equation}
and give interpretation for each term in Eq.~\ref{flapping2}.  The dimensionless geometrical prefactor, $L/I^{1/4}$, is apparent.  That constant measures compares the length of the flapper with its moment of inertia and is this a measure of the flapper slenderness. For long, thin flappers, this quantity is always larger than 1. For the type of flappers considered in our experiments, we have $6 < L/I^{1/4}< 35$. The dimensionless number in the second parenthesis of Eq.~\ref{flapping2} is a pseudo-Weissenberg number comparing the viscous stresses in the fluid with the elastic stresses in the beam. For all the flappers and tip speeds consider in our study, $V_{\rm tip}\mu/(L E)$ is always smaller than $10^{-4}$. The combination of values of these two dimensionless groups results in $F$ being of O(1).

\begin{figure}[t]
\centering
\includegraphics[width=0.6\textwidth]{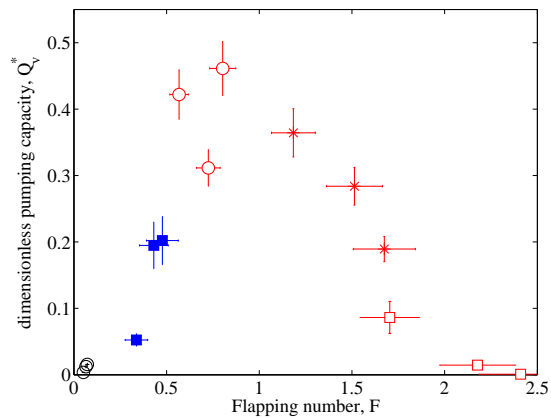}
\caption{Dimensionless pumping capacity, $Q_{v}^{*}$, as a function of the flapping number, $F$. The symbols labeling each data set are shown in Table 1. } \label{pumpflapping}
\end{figure}

In Fig.~\ref{pumpflapping} we plot the  dimensionless pumping capacity, $Q_v^*$,  as a function of $F$ for all
our experiments. All data collapse on a single master curve showing a maximum dimensionless pumping  for $F\approx1$, similarly to what  has been observed in the case of   flexible filaments actuated by an external force or moments \cite{Wiggins1998,Lowe2003,Dreyfus2005,yu06,lauga07_pre,Onshun2011}. No net flow is being pumped in the limit of very rigid or small flapper ($F\ll 1$) nor in the very long or flexible limit ($F\gg 1$).  It is notable that, under the optimal pumping conditions,  the dimensionless pumping capacity is close to 1/2. The net flow pumped during each period of oscillation is therefore about half of the flow periodically transported from one side of the tank to the next and indicates a high kinematic efficiency of the vertical pump proposed here.

\section{Conclusion}\label{conclusion}

Inspired by flexible actuators evolved by Nature  in the case of swimming microorganisms,  we have studied  experimentally the flow induced by a flexible flapper actuated periodically. The breaking of shape symmetry induced by the elasticity in the flapper allowed the creation of non-reciprocal  kinematics, and a net time-averaged flow.  We carried out detailed measurements to understand the physical origin of the net flow and showed that  flexibility induces a right-left asymmetry in the  main vortical structure induced by flapping, resulting in a net fluid upward fluid jet.  Our measurements also allowed us to demonstrate that the pumping capacity of the flapper was  maximum when the flapper length was on the same order as the elastohydrodynamic penetration length. Our setup could be used as a new mixing strategy in a viscous environment without the need of elaborate geometries or oscillating strategies. Specifically, the recirculation loops induced by our flapper would allow stirring of different layers of fluids aligned in the direction perpendicular to the flapper.  However, a more in depth study is needed to assess the mixing capacity of the flow produced by the flexible flapper. In particular, some of the ideas discussed in the mixing literature by laminar jets (see for instance \cite{rossi2012}) could be used to evaluate the mixing potential of the current setup. We hope that the proof-of-principle study carried out in this paper  will encourage further work in this direction.

\section{Acknowledgements}
This research was funded in part by the National Science Foundation (grant CBET-
0746285 to E. L.), PAPIIT-UNAM program (IN101312 to R.Z.) and the UC MEXUS-CONACYT program.
J.R. Velez-Cordero acknowledges the postdoctoral support of CONACyT-M\'exico and UC-MEXUS. R. Zenit is grateful to the M. Moshinsky foundation for its support.

\end{document}